# An Ontology-based System for Cloud Infrastructure Services' Discovery

Miranda Zhang[1, 2], Rajiv Ranjan[1], Armin Haller[1], Dimitrios Georgakopoulos[1], Michael Menzel[3], Surya Nepal[1]

[1] Information Engineering Laboratory, CSIRO ICT Centre
{miranda.zhang, rajiv.ranjan, armin.haller, dimitrios.georgakopoulos, surya.nepal}@csiro.au
[2] Research School of Computer Science, ANU
miranda.zhang@anu.edu.au
[3] Karlsruhe Institute of Technology, Karlsruhe, Germany
menzel@fzi.de

*Abstract*— **The Cloud infrastructure services landscape advances steadily leaving users in the agony of choice. As a result, Cloud service identification and discovery remains a hard problem due to different service descriptions, non-standardised naming conventions and heterogeneous types and features of Cloud services. In this paper, we present an OWL-based ontology, the Cloud Computing Ontology (CoCoOn) that defines functional and non-functional concepts, attributes and relations of infrastructure services. We also present a system, CloudRecommender-that implements our domain ontology in a relational model. The system uses regular expressions and SQL for matching user requests to service descriptions. We briefly describe the architecture of the CloudRecommender system, and demonstrate its effectiveness and scalability through a service configuration selection experiment based on a set of prominent Cloud providers' descriptions including Amazon, Azure, and GoGrid.**

## I. Motivation

The emergence of Cloud computing [1] over the past five years is potentially one of the breakthrough advances in the history of computing. The Cloud computing paradigm is shifting computing from in-house managed hardware and software resources to virtualized Cloud-hosted services. Cloud computing assembles large networks of virtualized services: hardware resources (CPU, storage, and network) and software resources (e.g., web server, databases, message queuing systems, monitoring systems.). Cloud services can be abstracted into three layers: Software as a Service (SaaS), Platform as a Service (PaaS), and Infrastructure as a Service (IaaS). Hardware and software resources form the basis for delivering IaaS and PaaS. The top layer focuses on application services (SaaS) by making use of services provided by the lower layers. PaaS/SaaS services are often developed and provided by third party service providers who are different from the IaaS providers. In this paper, we focus on IaaS that is the underpinning layer on which the PaaS services are hosted for creating SaaS applications.

From a service discovery point of view, the selection process on the IaaS layer is based on a finite set of functional and non-functional configuration properties (e.g. CPU type, memory size, costs, regional availability) that are potentially met by multiple providers. Similarly, there is a service discovery problem associated with the SaaS and PaaS offerings. However, we are not considering these issues in this paper.

IaaS providers include Amazon Web Services (AWS), Microsoft Azure, Rackspace, GoGrid, and others. They give users the option to deploy their application over a pool of virtually infinite services with practically no capital investment and with modest operating costs proportional to the actual use. Elasticity, cost benefits and abundance of resources motivate many organizations to migrate their enterprise applications to the Cloud. Although Cloud offers the opportunity to focus on revenue growth and innovation, decision makers (e.g., CIOs, scientists, developers, engineers, etc.) are faced with the complexity of choosing the right service delivery model for composite application and infrastructure across private, public, and hybrid Clouds.

Existing approaches which help a user to compare and select infrastructure services in Cloud computing involve manually reading the provider documentation for finding out services that are most suitable for hosting an application. This problem is further aggravated by the use of non-standardized naming conventions used by Cloud providers. For example, Amazon refers to compute services as EC2 Compute Unit, while GoGrid refers to the same as Cloud Servers. Furthermore, Cloud providers typically publish their service descriptions, pricing policies and Service-Level-Agreement (SLA) rules on their websites in various formats. The relevant information may be updated without prior notice to the users. Hence, it is not an easy task to manually obtain and compare service configurations from Cloud providers' websites and documentations (which are the only sources of information).

Although popular search engines (e.g., Google, Bing, etc) can point users to these provider web sites (blogs, wikis, etc.) that describe IaaS service offerings, they are not designed to compare and reason about the relations among the different types of Cloud services and their configurations. Service description models and discovery mechanisms for determining the similarity among Cloud infrastructure services are needed to aid the user in the discovery and selection of the most cost effective infrastructure service meeting the user's functional and non-functional requirements.

In order to address these aforementioned problems, we present a semi-automated, extensible, and ontology-based approach to infrastructure service discovery and selection and its implementation in the CloudRecommender system. We identify and formalize the domain knowledge of multiple configurations of infrastructure services. The core idea is to formally capture the domain knowledge of services using semantic Web languages like the Resource Description Framework (RDF) and the Web Ontology Language (OWL). The contributions of this paper are as the following:

- Identification of the most important concepts and relations of functional and non-functional configuration parameters of infrastructure services and their definition in an ontology;
- Modelling of service descriptions published by Cloud providers according to the developed ontology. By doing so, we validate the expressiveness of ontology against the most commonly available infrastructure services including Amazon, Microsoft Azure, GoGrid, etc.
- An implementation of a design support system, CloudRecommender, based on our ontological model for the selection of infrastructure Cloud service configurations using transactional SQL semantics, procedures and views. The benefits to users of CloudRecommender include, for example, the ability to estimate costs, compute cost savings across multiple providers with possible tradeoffs and aid in the selection of Cloud services.

The remainder of the paper is structured as follows. A discussion on our formal domain model for Cloud infrastructure services is presented in Section 2. Details on the proposed Cloud selection approach and the CloudRecommender system are given in Section 3. A review of related work is provided in Section 4 before we conclude in Section 5.

## II. CLOUD COMPUTING ONTOLOGY

The Cloud Computing Ontology (CoCoOn) defines the domain model of the IaaS layer. This ontology facilitates the description of Cloud infrastructure services; and through mappings from provider descriptions, facilitates the discovery of infrastructure services based on their functionality and Quality of Service (QoS) parameters. The ontology is defined in the OWL [12] and can be found at: http://w3c.org.au/cocoon.owl. To describe specific aspects of Cloud computing, established domain classifications have been used as a guiding reference [9, 11]. For the layering of the ontology on top of Web service models, it builds upon standard semantic Web service ontologies i.e., OWL-S [10] and WSMO [13]. Consequently, modellers can use the grounding model and process model of OWL-S in combination with the presented Cloud computing ontology to succinctly express common infrastructure Cloud services. We mapped the most prominent set of infrastructure services (i.e. Amazon, Azure, GoGrid, Rackspace, etc.) to CoCoOn. All common metadata fields in the ontology including Organisation, Author, First Name etc. are referenced through standard Web Ontologies (i.e. FOAF[1] and Dublin Core[2]).

The Cloud computing ontology consists of two parts: functional Cloud service configurations information parameters; and non-functional service configuration parameters. In the following subsections, we detail on these two parts. We also present parts of the ontology in a visual form produced by the Cmap Ontology Editor tool [14].

*1) Functional Cloud service configuration parameters*

The main concept to describe functional Cloud service configurations in CoCoOn is a *CloudResource* that can be of one of the three types: Infrastructure-as-a-Service (*IaaS*), Platform-as-a-Service (*PaaS*) or Software-as-a-Service (*SaaS*). For the current implementation of the CloudRecommender system, we have defined the Cloud infrastructure layer (IaaS), providing concepts and relations that are fundamental to the other higher-level layers. In future work, we will extend the ontology to cover both PaaS and SaaS layers.

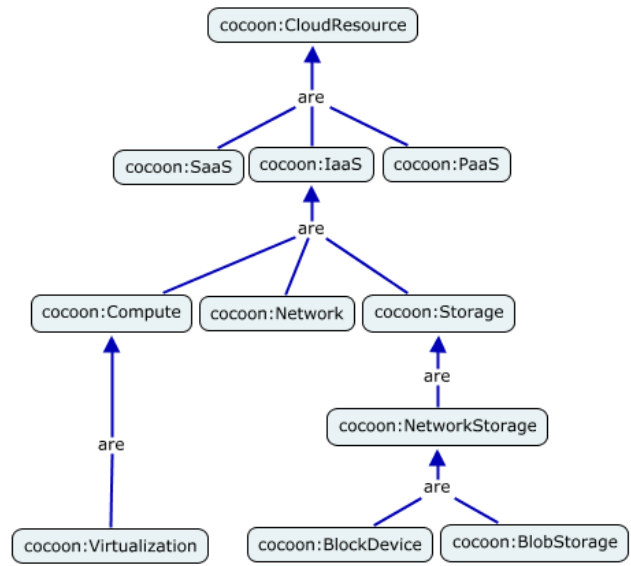

FIGURE 1: TOP CONCEPTS IN THE IAAS LAYER

Cloud services in the IaaS layer can be categorised into: *Compute*, *Network*, and *Storage* services (see Fig. 1). *Compute* is the main concept for infrastructure services, whereas *Network* and *Storage* are usually attached to a *Compute* service (with exceptions, for example *NetworkStorage*, see below).

---
[1] See http://xmlns.com/foaf/spec/

[2] See http://purl.org/dc/elements/1.1

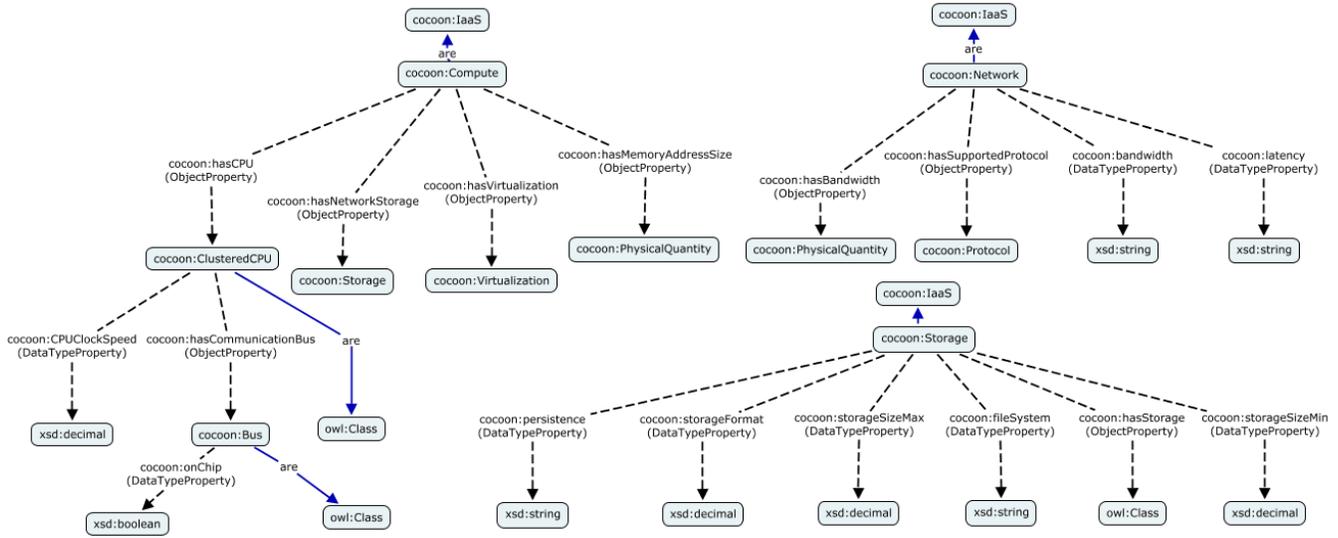

FIGURE 2: SUBCLASSES AND PROPERTIES FOR THE COMPUTE, STORAGE AND NETWORK CLASS

The *Compute* class (see Fig. 2) has the following object properties, *hasVirtualization*, *hasCPU*, *hasMemoryAddressSize* and *hasNetworkStorage*. The *hasCPU* property links a *Compute* unit to one or many processors which can be of type *CPU* or *ClusteredCPU*. A *Compute* object can be linked to a *Storage* object by using the top level object property *hasStorage*.

There are two different *Storage* types for a *CloudResource*: *LocalStorage* attached to a *CPU* with the *hasLocalStorage* property and *NetworkStorage* attached to a *Compute* instance with the *hasNetworkStorage* property. The *hasNetworkStorage* is an *owl:inverseOf* property of the *isAttached* property which can be used to define that a *Storage* resource is attached to a *Compute* resource. There is also an important distinction to be made between *Storage* resources that are attached to a *Compute* resource and *Storage* resources that can be attached. The latter is modeled with the *isAttachable* object property and its inverse property *hasAttachable*. These relations are important for the discovery of infrastructure services based on a user requirement. For example, in the case of Amazon, we can model that a *BlockStorage* with a *StorageSizeMin* of 1GB and a *StorageSizeMax* of 1TB can be attached to any *EC2 Compute* resource instance i.e., *Standard*, *Micro*, *High-Memory*, *High-CPUCluster*, *ComputeCluster*, *GPUHigh-I/O*. Consequently, if a user searches for a specific *Compute* instance with, for example, 5GB persistent storage, the relevant *EC2 Compute* resource and an Amazon *BlockStorage* will be returned (possibly among others). That is, because the *isAttached* relation in the user request can be matched with the definition of the Amazon EC2 unit with a *BlockStorage* defined to be *isAttachable*.

A *Network* resource can be described with the *hasBandwidth* and *hasProtocol* properties. Similarly, to how *Storage* resources are attached to *Compute* resource, we distinguish between the *hasSupportedNetwork* and *hasNetwork* property to either express that the specific network types can be used with a *Compute* resource or that they are in fact used.

*2) Non-Functional Cloud service configuration parameters*

For non-functional Cloud service configuration parameters we distinguish between non-functional properties and QoS attributes. The first are properties of Cloud resources that are known at design time, for example, *PriceStorage*, *Provider*, *DeploymentModel*, whereas QoS attributes can only be recorded after at least one execution cycle of a Cloud service, for example, *DiskReadOperations*, *NetworkIn*, *NetworkOut* etc. For QoS attributes, we distinguish *MeasurableAttributes* like the ones above and *UnmeasurableAttributes* like *Durability* or *Performance*.

The QoS attributes define a taxonomy of *Attributes* and *Metrics*, i.e. two trees formed using the *rdfs:subClassOf* relation where a *ConfigurationParameter*, for example, *PriceStorage*, *PriceCompute*, *PriceDataTransferIn* (Out) etc. and a *Metric*, for example, *ProbabilityOfFailureOnDemand*, *TransactionalThroughput*, are used in combination to define non-functional properties (e.g. *Performance*, *Cost*, etc.). The resulting ontology is a (complex) directed graph where, for example, the Property *hasMetric* (and its inverse *isMetricOf*) is the basic link between *ConfigurationParameters* and *Metric* trees. For the QoS metrics, we used existing QoS ontologies [16] as a reference whereas for the *ConfigurationParameters* concepts the ontology defines its independent taxonomy, but refers to external ontologies for existing definitions (e.g. QUDT [3]). Each configuration parameter (compare Table I) has a *Name* and a *Metric* (qualitative or quantitative). The *Metric* itself has a *UnitOfMeasurement* and a *Value*. The type of configuration determines the nature of a service by means of setting a minimum, maximum, or capacity limit, or meeting a certain value. For example, the *hasMemory* configuration parameter of a Compute service can be set to have a Value of 2 and a *UnitOfMeasurement* of GB.

### III. A SYSTEM FOR CLOUD SERVICE SELECTION

We propose an approach and a system for Cloud service configuration selection called CloudRecommender. For our CloudRecommender service, we implemented the Cloud Service Ontology as a relational model and the Cloud QoS ontology as configuration information as structured data (entities) which we query using SQL. The choice of a relational model and SQL as query language was made because of the convenience SQL procedures offers us in regards to defining templates for a given widget type (see below). We use stored procedures to create temporary tables and to concatenate parameters to dynamically generate queries based on the user input. As a future work, we will migrate the infrastructure services definitions to an RDF database and use, for example, SPIN templates to encode our procedures in SPARQL.

We collected service configuration information from a number of public Cloud providers (e.g., Windows Azure, Amazon, GoGrid, RackSpace, Nirvanix, Ninefold, SoftLayer, AT and T Synaptic, Cloud Central, etc.) to demonstrate the generic nature of the domain model with respect to capturing heterogeneous configuration (see Table II) information of infrastructure services. The CloudRecommender system architecture (shown in Fig. 3) consists of three layers: the configuration management layer, the application logic layer and the User interface (widget) layer. Details of each layer will be explained in the following sub-sections.

#### A. Infrastructure service configuration repository

The system includes a repository of available infrastructure services from different providers including compute, storage and network services. These infrastructure services have very different configurations and pricing models. Distinct and ambiguous terminologies are often used to describe similar configurations, for example different units of measurements are used for similar metrics. We performed unit conversions during instantiation of concepts to simplify the discovery process. For example, an Amazon EC2 Micro Instance has 613 MB of memory which is converted to approximately 0.599 GB. Another example is the CPU clock speed. Amazon refers to it as "ECUs". From their documentation [7]: *"One EC2 Compute Unit provides the equivalent CPU capacity of a 1.0-1.2 GHz 2007 Opteron or 2007 Xeon processor. This is also the equivalent to an early-2006 1.7 GHz Xeon processor referenced in our original documentation"*.

---

[3] See http://www.qudt.org

Another example of disparity between different Cloud providers is the price model of "on Demand instances". GoGrid's plan, although having a similar concept to Amazon's On Demand and Reserved Instance, gives very little importance to what type or how many of compute services a user is deploying. GoGrid charges users based on what they call RAM hours – 1 GB RAM compute service deployed for 1 hour consumes 1 RAM Hour. A 2 GB RAM compute service deployed for 1 hour consumes 2 RAM Hour. It is worthwhile mentioning that only Azure clearly states that one month is considered to have 31 days. This is important as the key advantage of the fine grained pay-as-you-go price model which, for example, should charge a user the same when they use 2GB for half a month or 1 GB for a whole month. Other vendors merely give a GB-month price without clarifying how short term usage is handled. It is neither reflected in their usage calculator. We chose 31 days as default value in calculation.

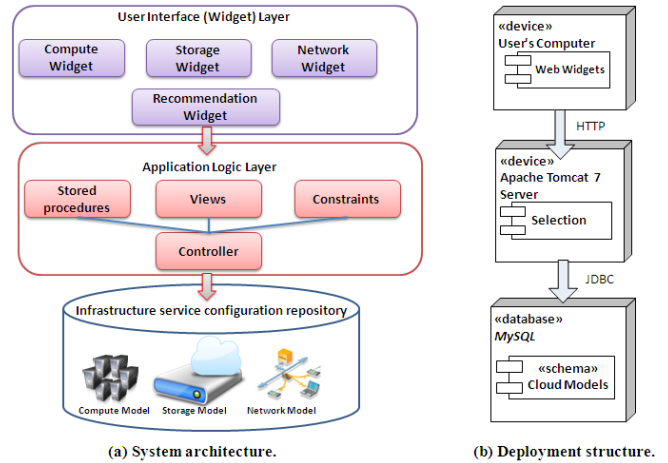

FIGURE 3: CLOUDRECOMMENDER SYSTEM ARCHITECTURE.

Regarding storage services, providers charge for every operation that an application program or user undertakes. These operations are effected on storage services via REpresentational State Transfer (RESTful) application programming interfaces (APIs) or Simple Object Access Protocol (SOAP) API. Cloud providers refer to the same set of operations with different names, for example Azure refers to storage service operations as transactions. Nevertheless, the operations are categorized into upload and download categories as shown in Table III. Red means an access fee is charged; green means the service is free; and yellow means access fees are not specified, and can usually be treated as green/free of charge. To facilitate our calculation of similar and equivalent requests across multiple providers, we analyzed and pre-processed the price data, recorded it in our domain model and used a homogenized value in the repository (configuration management layer). For example, Windows Azure Storage charges a flat price per transaction. It is considered as transaction whenever there is a "touch" operation, i.e. Create, Read, Update, Delete (CRUD) operation over the RESTful service interface, on any component (Blobs, Tables or Queues) of Windows Azure Storage.

For providers that offer different regional prices, we store the location information in the price table. If multiple regions have the same price, we choose to combine them. In our current implementation, any changes to existing configurations (such as updating memory size, storage provision etc.) of services can be done by executing customized update SQL queries. We also use customized crawlers to update provider information's periodically. However, as a future work, we will provide a RESTful interface and widget that can be used for automatic configuration updates.

*B. Application Logic Layer*

The request for service selection in CloudRecommender is expressed as SQL queries. The selection process supports an application logic that builds upon views and stored procedures.

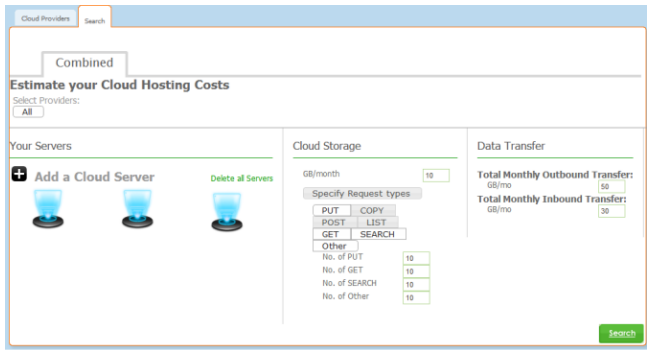

FIGURE 4: SCREEN SHOT OF COMPUTE, STORAGE, AND NETWORK WIDGETS.

*C. User Interface Layer*

This layer features a rich set of user-interfaces (see Fig. 4) that further simplify the selection of configuration parameters related to Cloud services. This layer encapsulates the user interface components in the form of four principle widgets including: Compute, Storage, Network, and Recommendation. The selection of basic configuration parameters related to compute services including their RAM capacity, cores, and location can be facilitated through the Compute widget. It also allows users to search compute services by using regular expressions, sort by a specific column etc. Using the Compute widget, users can choose which columns to display and rearrange their order as well. The Storage widget allows users to define configuration parameters such as storage size and request types (e.g., GET, PUT, POST, COPY etc.). Service configuration parameters, such as the size of incoming data transfer and outgoing data transfer can be issued via the Network widget. Users have the option to select single service types as well as bundled (combined search) services driven by use cases. The selection results are displayed and can be browsed via the Recommendation widget (not shown in Fig. 4).

IV. RELATED WORK

Currently, there are 3 common approaches for Cloud services identification/publication: 1) manually maintain directories by categorizing submitted or collected information about Cloud services and providers, an example of such kind is Universal Description, Discovery and Integration (UDDI), which has failed to gain wide adoption; 2) use of web crawlers for automatically creating service listings; and 3) combining both of the aforementioned approaches, e.g. using manually-submitted URIs as seeds to generate indexes. The first approach is the only feasible solution at the moment. Some of the recent research such as [6] has focused on Cloud storage and network service (IaaS level) representation using XML. However, the proposed schema does not comply with or take into account any standardization efforts proposed as ontologies on the semantic web.

Notably branded calculators are available from individual Cloud providers, such as Amazon [[4]], Azure [[5]], and GoGrid, for service leasing cost calculation. However, it is not easy for users to generalize their requirements to fit different service offers (with various quota and limitations) let alone computing and comparing costs. All of the aforementioned calculators name and represent service configurations differently, hence making the task of unified service selection and comparison impossible.

Although the authors in [15] present a taxonomy (ontology) to classify Cloud services across IaaS, PaaS, and SaaS layers, they fail to capture low-level configuration information of services and their dependencies across layers. Furthermore, their taxonomy does not include concepts and relationships pertaining to QoS configuration of services.

Overall, the proposed generic ontology and its implementation in relational CloudRecommender system is preferable over hard coding the sorting and selection algorithm as it allows us to take the advantage of optimized SQL operations (e.g. select and join).

| Service | Configurations Parameters | Range/possible values |
|---|---|---|
| Compute | Core | >= 1 |
| | CPUClockSpeed | > 0 |
| | hasMemory | > 0 |
| | hasCapacity | >= 0 |
| | Location | North America, South America, Africa, Europe, Asia, Australia |
| | CostPerPeriod | >= 0 |
| | PeriodLength | > 0 |
| | CostOverLimit | >= 0 |
| | PlanType | Pay As You Go, Prepaid |
| Storage | StorageSizeMin | >= 0 |
| | StorageSizeMax | > 0 |
| | CostPerPeriod (e.g. Period = Month) (e.g. UnitOfMeasurement = GB) | >= 0 |
| | Location | North America, South America, Africa, Europe, Asia, Australia |
| | RequestType | put, copy, post, list, get, delete, search |
| | CostPerRequest | >= 0 |
| | PlanType | Pay As You Go, Prepaid, Reduced Redundancy |
| Network | CostDataTransferIn | >= 0 |
| | CostDataTransferOut | > 0 |

TABLE I. INFRASTRUCTURE SERVICE TYPES AND THEIR CONFIGURATIONS.

| Provider | Compute Terminology | Pay As You Go Unit | Other Plans* | Storage Terminology | Pay As You Go Unit | Other Plans* | Trail Period or Value |
|---|---|---|---|---|---|---|---|
| Windows Azure | Virtual Server | /hr | 1 | Azure Storage | /GB month | 1 | 90 day |
| Amazon | EC2 Instance | /hr | 2 | S3 | /GB month | 2 | 1 year |
| GoGrid | Cloud Servers | /RAM hr | 1 | Cloud Storage | /GB month | | |
| RackSpace | Cloud Servers | /RAM hr | | Cloud Files | /GB month | | |
| Nirvanix | | | | CSN | /GB month | | |
| Ninefold | Virtual Server | /hr | | Cloud Storage | /GB month | 1 | 50 AUD |
| SoftLayer | Cloud Servers | /hr | 1 | Object Storage | /GB | | |
| AT and T Synaptic | Compute as a Service | vCPU per hour + /RAM hr | | Storage as a Service | /GB month | | |
| Cloudcentral | Cloud Servers | /hr | | | | | |
| * Monthly/Quarterly/Yearly Plan, Reserve and Bidding Price Option | | | | | | | |

TABLE II. DEPICTION OF CONFIGURATION HETEROGENEITIES IN COMPUTE AND STORAGE SERVICES ACROSS PROVIDERS. (RED) BLANK CELLS IN THE TABLE MEAN THAT A CONFIGURATION PARAMETER IS NOT SUPPORTED. SOME PROVIDERS OFFER THEIR SERVICES UNDER A DIFFERENT PRICING SCHEME THAN PAY-AS-YOU-GO. IN TABLE II WE REFER TO THESE SCHEMES AS OTHER PLANS (E.G. AMAZON REDUCED REDUNDANCY, RESERVED PRICE PLANS, GOGRID PRE-PAID PLANS)

| Provider | Storage | Requests | | |
|---|---|---|---|---|
| | | Upload | Download | Other |
| Windows Azure | Azure Storage | storage transactions | storage transactions | |
| Amazon | S3 | PUT, COPY, POST, or LIST Requests | GET and all other Requests | Delete |
| GoGrid | Cloud Storage | Transfer protocols such as SCP, SAMBA/CIFS, and RSYNC | | |
| RackSpace | Cloud Files | PUT, POST, LIST Requests | HEAD, GET, DELETE Requests | |
| Nirvanix | CSN | | Search | |
| Ninefold | Cloud Storage | GET, PUT, POST, COPY, LIST and all other transactions | | |
| SoftLayer | Object Storage | Not Specified/Unknow | | |
| AT and T Synaptic | Storage as a Service | Not Specified/Unknow | | |

TABLE III. DEPICTION OF CONFIGURATION HETEROGENEITIES IN REQUEST TYPES ACROSS STORAGE SERVICES.

## V. CONCLUSION AND FUTURE WORK

In conclusion, in this paper, we have proposed ontology for classifying and representing the configuration information related to Cloud-based IaaS services including compute, storage, and network. The proposed ontology is comprehensive as it can not only capture static configuration but also dynamic QoS configuration on the IaaS layer. We also presented the implementation of the ontology in the CloudRecommender system. The paper will help readers in clearly understanding the core IaaS-level Cloud computing concepts and inter-relationship between different service types. This in turn may lead to a harmonization of research efforts and more inter-operable Cloud technologies and services at the IaaS layer.

In future work, we intend to extend the ontology with the capability to store PaaS and SaaS configurations. Moreover, we would also like to extend our ontology to capture the dependency of services across the layers. For example, investigating concepts and relationships for identifying the dependencies between compute service (IaaS) configurations and the type of appliances (PaaS) that can be deployed over it. For instance, before mapping a MySQL database appliance (PaaS) to a Amazon EC2 compute service (IaaS), one needs to consider whether they are compatible in terms of virtualization format. Another avenue that we would like to explore is how to aggregate QoS configurations across the IaaS, PaaS, and SaaS layers for different application deployment scenarios (e.g., multimedia, eResearch, and enterprise applications).